# Cultural Evolution as Distributed Computation

LIANE GABORA
University of British Columbia

## Abstract
The speed and transformative power of human cultural evolution is evident from the change it has wrought on our planet. This chapter proposes a human computation program aimed at (1) distinguishing algorithmic from non-algorithmic components of cultural evolution, (2) computationally modeling the algorithmic components, and amassing human solutions to the non-algorithmic (generally, creative) components, and (3) combining them to develop human-machine hybrids with previously unforeseen computational power that can be used to solve real problems. Drawing on recent insights into the origins of evolutionary processes from biology and complexity theory, human minds are modeled as self-organizing, interacting, autopoietic networks that evolve through a Lamarckian (non-Darwinian) process of communal exchange. Existing computational models as well as directions for future research are discussed.

## Introduction

The origin of life brought about unprecedented change to our planet; new forms emerged creating niches that paved the way for more complex forms, completely transforming the lands, skies, and oceans. But if biological evolution is effective at bringing about adaptive change, human cultural evolution is arguably even more effective, and faster. Cultural change doesn't take generations; it works at the speed of thought, capitalizing on the strategic, intuitive creative abilities of the human mind.

This chapter outlines current and potential future steps toward the development of a human computation program inspired by the speed and effectiveness of how culture evolves. The overarching goal of the kind of research program outlined in this chapter is to develop a scientific framework for cultural evolution by abstracting its algorithmic structure, use this algorithmic structure to develop human-machine hybrid structures with previously unforeseen computational power, and to apply it to solving real problems. The proposed approach can be thought of as a "repeatable method" or "design pattern" for fostering cultural emergence, defined by specific computational methods for modeling interactions at the conceptual level, the individual level, and the social level, and their application to the accumulation of adaptive, open-ended cultural novelty.

## Two Approaches to a Scientific Framework for Culture

Cultural evolution entails the generation and transmission of novel behavior and artifacts within a social group, both vertically from one generation to another, and horizontally amongst members of a generation. Like biological evolution, it relies on mechanisms for both introducing variation and preserving fit variants. Elements of culture adapt, diversify, and become more complex over time, and exhibit phenomena observed in biological evolution, such as niches, drift, epistasis, and punctuated equilibrium (Bentley, Hahn & Shennan, 2004; Durham, 1991; Gabora, 1995). However, we lack a precise understanding of how culture evolves.

We begin by summarizing two approaches that have been taken to developing a formal understanding of the process by which culture evolves: Darwinian approaches, and Communal Exchange approaches.

*Darwinian Approaches*
Dawkins' (1975) proposal that culture evolves through reiterated variation and selection inspired formal Darwinian models of cultural evolution (Boyd & Richerson, 1985, 2005; O'Brien, M.J., & Lyman Cavalli-Sforza & Feldman, 1981; Henrich & Boyd, 1998, 2002). It also inspired some archaeologists to apply methods designed for documenting the evolution of biological organisms to chart the historical evolution of artifacts (*e.g.*, O'Brien, & Lyman, 2000; Shennan, 2008). Aside from the questionable assumptions underlying this approach (Atran, 2001; Fracchia & Lewontin, 1999; Gabora, 1997, 1998, 1999, 2001, 2004, 2006a, 2008, 2011; Skoyles, 2008; Temkin & Eldredge, 2007), it aims to model how cultural variants spread, not how they come into existence, strategically building on and opening up new niches for one another.

Holland (1975) elucidated the algorithmic structure of natural selection, resulting in the *genetic algorithm* (GA), and subsequently genetic programming (GP) (Koza, 1993), optimization tools with diverse applications to everything from scheduling tasks (Hou, Ansari, & Ren, 1994) to pipeline design (Goldberg, Kuo, 1987) to music and art (Bentley & Corne, 2002; DiPaola & Gabora, 2009). The term *cultural algorithm* has referred to a GA that includes a 'belief space' used to prune the search space (Reynolds, 1994), not an algorithm inspired by how culture itself evolves. GAs are effective for multi-constraint problems with complex fitness landscapes, but would not do well on problems that require reformulating or *restructuring* the problem from another perspective. GAs are *breadth-first* (generate *many* solutions *randomly*, and some by chance may be effective), whereas cultural evolution, which relies on cognitive processes such as learning, is *depth-first* (generate *few* solutions making use of strategic analysis or spontaneous associations, either intentional or unintentional).

*Communal Exchange*
Mounting evidence suggests that a non-Darwinian framework is appropriate for, not just cultural evolution, but the earliest stages of organic life itself (Gabora, 2006; Kauffman, 1993; Vetsigian, Woese, & Goldenfeld, 2006; Williams & Frausto da Silva, 2003), and aspects of modern microbial life (Woese 2002). There is widespread support for the hypothesis that the earliest protocells were self-organized autocatalytic networks that evolved (albeit haphazardly) through a non-Darwinian process involving horizontal transfer of innovation protocols, referred to as *communal exchange* (Vetsigian, Woese, & Goldenfeld, 2006). Communal exchange differs substantially from natural selection. Acquired change is retained, and information is transmitted communally, not by way of a self-assembly code from parent to offspring. Formal methods for modeling reaction networks can be used to investigate the feasibility of the emergence of the kind of self-sustaining structure that could evolve through communal exchange.

It has been suggested that the basic unit of cultural evolution is, not an autocatalytic network per se, but an associative network that is (like an autocatalytic network) *autopoietic, i.e.*, the whole emerges through interactions amongst the parts (Gabora, 1999, 2004). A communal exchange based computational model of cultural evolution has been developed (Gabora, 1995, 2008a,b). EVOC (for EVOlution of Culture) consists of neural network based agents that invent new actions and imitate actions performed by neighbors. The assemblage of ideas changes over time not because some replicate at the expense of others, as in natural selection, but because they transform through inventive and social processes. Agents can make generalizations concerning what kinds of actions are fittest, and use this acquired knowledge to modify ideas for actions between transmission events. EVOC exhibits typical evolutionary patterns, *e.g.*, cumulative increase in fitness and complexity of cultural elements over time, and an increase in diversity as the space of possibilities is explored, followed by a decrease as agents find and converge on the fittest possibilities. EVOC has been used to model how the mean fitness and diversity of cultural elements is affected by factors such as leadership, population size and density, borders that affect transmission between populations, and the proportion and distribution of creators (who acquire new ideas primarily by inventing them) versus imitators (who acquire new ideas primarily by copying their neighbors) (Gabora, 1995, 2008a,b; Gabora, & Firouzi, 2012; Gabora & Leijnen, 2009; Leijnen & Gabora, 2010).

A communal exchange inspired method for organizing artifacts into historical lineages has also been developed. *Worldview Evolution*, or WE for short, uses both superficial (*e.g.*, 'beveled edge') and abstract (*e.g.*, 'object is thrown') attributes, as well as analogical transfer (*e.g.*, of 'handle' from knife to cup) and complementarity (*e.g.*, bow and arrow) (Gabora, Leijnen, Veloz, & Lipo, 2011). It represents objects not in terms of a convenient list of discrete measurable attributes, but in terms of how they are actually conceptualized, as a network of interrelated properties, using a *perspective* parameter that can be weighted differently according to their relative importance. Preliminary analyses show that the conceptual network approach can recover previously unacknowledged patterns of historical relationship that are more congruent with geographical distribution and temporal data than is obtained with an alternative cladistic approach that is based on the assumption that cultural evolution, like biological evolution, is Darwinian.

These two computational models, EVOC and WE, show that a communal exchange approach to cultural evolution is computationally tractable. However such models will not begin to approach the open-ended ingenuity and complexity of human cultural evolution until they incorporate certain features of the cognitive process by which cultural novelty is generated.

## The Generation of Cultural Novelty

We said that cultural evolution is a depth-first evolution strategy. A depth-first evolution strategy entails processes that adaptively bias the generation of novelty. A number of key, interrelated processes have been identified that, in addition to learning, accomplish this in cultural evolution. We now look briefly at some of these processes, as well as efforts to model them.

### *Recursive Recall and Restorative Restructuring*

*Recursive recall* (RR) is the capacity for one thought to trigger another, enabling progressive modification of an idea. Donald's (1991) hypothesis that cultural evolution was made possible by onset of the capacity for RR has been tested using EVOC (Gabora & Saberi, 2011; Gabora & DiPaola, 2012). A comparison was made of runs in which agents were limited to single-step actions to runs in which they could recursively operate on ideas, and chain them together, resulting in more complex actions. While RR and no-RR runs both converged on optimal actions, without RR this set was static, but with RR it was in constant flux as ever-fitter actions were found. In RR runs there was no ceiling on mean fitness of actions, and RR enhanced the benefits of learning.

Although these findings support Donald's hypothesis, the novel actions generated with RR were predictable. They did not open up new cultural niches in the sense that, for example, the invention of cars created niches for the invention of things like seatbelts and stoplights. EVOC in its current form could not solve *insight problems*, which require restructuring the solution space (Boden, 1990; Kaplan & Simon, 1990, Ohlsson, 1992). Restructuring can be viewed as a form of RR that entails looking at the problem from a new context or perspective, and that this is driven by the mind's *self-organizing, restorative* capacity.

### *Contextual Focus (CF) and Divergent versus Associative Thought*

It has been proposed that restorative restructuring is aided by *contextual focus* (CF): the capacity to spontaneously and temporarily shift to a more divergent mode of thought (Gabora, 2003). Divergent thought entails an increase in activation of the associates of a given item (Runco, 2010). Thus for example, given the item TABLE, in a convergent mode of thought you might call to mind accessible associates such as CHAIR, but in a divergent mode of thought you might also call to mind more unusual associates such as PICNIC or MULTIPLICATION TABLE. CF has been implemented in EVOC (the computational model of cultural evolution). Low fitness of ideas induces a temporary shift to a more divergent processing mode by increasing the 'reactivity', $\alpha$, which determines the degree to which a newly invented idea can differ from the idea on which it was based.

Current research on the architecture of memory suggests that creative thought is actually not divergent but associative, as illustrated in Figure 1 (Gabora, 2010; Gabora & Ranjan, 2013). While divergent thought refers to an increase in activation of *all* associates, associative thought increases only activation of those relevant to the context. Because memory is distributed and content-addressable, associations are forged by way of shared structure. In associative thought, items come together that, though perhaps seemingly different, *share properties or relations,* and are thus more likely than chance to be *relevant* to one another, perhaps in a previously unnoticed but useful way.

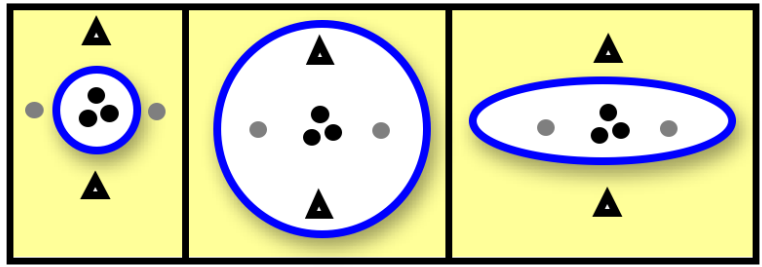

Figure 1. *Convergent thought* (left) activates key properties only, represented by black dots. *Divergent thought* (centre) activates not just key properties but also peripheral (less salient) properties, represented by both grey dots and black triangles. The grey dots represent peripheral properties that are relevant to the current context (goal or situation); the black triangles represent peripheral properties that are irrelevant to the current context. *Associative thought* (right) activates key properties and context-relevant peripheral properties.

A processing mode that is not just divergent but associative could be simulated in a model such as EVOC capitalizing on the ability to learn generalizations (*e.g.,* symmetrical movements tend to be fit) to constrain changes in $\alpha$. It would also be interesting to investigate the topological and dynamical properties of fitness landscapes for which divergent versus associative forms of CF is effective. CF is expected to be most beneficial for fitness landscapes that are rugged and subject to infrequent, abrupt change, with associative CF outperforming divergent CF.

*Concept Interaction*
Since creative processes such as restructuring involve putting concepts together in new contexts, a model of cultural evolution should be built upon a solid theory of concepts and *how they interact*. However, people use conjunctions and disjunctions of concepts in ways that violate the rules of classical logic; *i.e.*, concepts interact in ways that are non-compositional (Osherson & Smith, 1981; Hampton, 1987; Aerts, 2009; Aerts, Aerts, & Gabora, 2009; Aerts, Broekaert & Gabora, 2010; Kitto, Ram, Sitbon, & Bruza, 2011). This is true both with respect to properties (*e.g.*, although people do not rate 'talks' as a characteristic property of PET or BIRD, they rate it as characteristic of PET BIRD), and exemplar typicalities (*e.g.*, although people do not rate 'guppy' as a typical PET, nor a typical FISH, they rate it as a highly typical PET FISH). Because of this, concepts have been resistant to mathematical description.

This non-compositionality can be modeled using a generalization of the formalisms of quantum mechanics (QM) (Aerts & Gabora, 2005; Gabora & Aerts, 2002a,b; Kitto, Ramm, Sitbon, & Bruza, 2011). The reason for using the quantum formalism is that it allows us to describe the chameleon-like way in which concepts interact, spontaneously shifting their meanings depending on what other concepts are nearby or activated. The following formal exposition, though not essential for grasping the underlying concepts, is provided for the mathematically inclined reader. In QM, the state $|\psi\rangle$ of an entity is written as a linear superposition of a set of basis states $\{|f_i\rangle\}$ of a complex Hilbert space H. Hence $|\psi\rangle = \Sigma_i c_i |f_i\rangle$ where each complex number coefficient $c_i$ of the linear superposition represents the contribution of each component state $|f_i\rangle$ to the state $|\psi\rangle$. The square of the absolute value of each coefficient equals the weight of its component basis state with respect to the global state. The choice of basis states is determined by the observable to be measured. The basis states corresponding to this observable are called *eigenstates*. Upon measurement, the state of the entity *collapses* to one of the eigenstates. In the quantum inspired State COntext Property (SCOP) theory of concepts, the basis states represent states (instances or exemplars) of a concept, and the measurement is the context that causes a particular state to be evoked. SCOP is consistent with experimental concept data on concept combination (Aerts, 2009; Aerts, Aerts, & Gabora, 2009; Aerts, Broekaert, Gabora, & Veloz, 2012; Aerts, Gabora, & Sozzo, submitted; Hampton, 1987), and with findings that a compound's constituents are not just conjointly activated but bound together in a context-specific manner that takes relational structure into account (Gagné & Spalding, 2009). The model is being expanding to incorporate larger conceptual structures (Gabora & Aerts, 2009), and different modes of thought (Veloz, Gabora, Eyjolfson, & Aerts, 2011). This theoretical work is complemented by empirical studies aimed at establishing that (i) some concept combinations involve interference and entangled states, and (ii) creative products are external evidence of an internal self-organization process aimed at resolving dissonance and restoring equilibrium through the recursive actualization of potentiality (Gabora, 2011; Gabora, O'Connor, & Ranjan, 2012; Gabora & Saab, 2011).

**Harnessing the Computational Power of Cultural Evolution**
We have looked at some of the key milestones that have been crossed in the development of a scientific framework for how culture evolves. These milestones include a crude but functional computational model of cultural evolution, research into the cognitive mechanisms underlying the generation of cultural novelty, and preliminary efforts to computationally model these mechanisms. The rest of this chapter presents new, untested, yet-to-be-implemented ideas for how to go about harnessing the speed and power of cultural evolution in the development of a human computation research program.

*Computational Model of Restorative Restructuring*
A first step is to develop a model of problem restructuring using a "reaction network" inspired model that has as its basic unit, not catalytic molecules, but interacting concepts. There are various methods for going about this, for example using Concat, or Holographic Reduced Representations to computationally model the *convolution* or 'twisting together' of mental representation (Aerts, Czachor, & De Moor, 2009; Eliasmith & Thagard, 2001; Thagard & Stewart, 2011). Another promising route is to use a quantum-inspired theory of concepts such as SCOP that incorporates the notion of context-driven actualization of potential (Aerts & Gabora, 2005a,b; Gabora & Aerts, 2002a,b). A concept is defined in terms of (1) its set of states or exemplars S, each of which consists of a set L of relevant properties, (2) set M of contexts in which it may be relevant, (3) a function *n* that describes the applicability or *weight* of a certain property for a specific state and context, and (4) a function *μ* that describes the transition probability from one state to another under the influence of a particular context.

The procedure is best explained using an example, such as the idea of using a tire to make a swing, *i.e.*, the invention of a tire swing (from Gabora, Scott, & Kauffman, in press). The concept TIRE consists of the set S of states of TIRE, and in the context 'winter', TIRE might collapse to SNOW TIRE. Suppose that the network's initial conception of TIRE, represented by vector $|p\tilde{n}$ of length equal to 1, is a superposition of only two possibilities (Fig. 2). The possibility that the tire has sufficient tread to be *useful* is denote by unit vector $|u\tilde{n}$. The possibility that it should be discarded as *waste* is denoted by unit vector, $|w\tilde{n}$. Their relationship is given by the equation $|p\tilde{n} = a_0|u\tilde{n} + a_1|w\tilde{n}$, where $a_0$ and $a_1$ are the amplitudes of $|u\tilde{n}$ and $|w\tilde{n}$ respectively. If a tire us useful only for transportation, denoted $|t\tilde{n}$ then, $|u\tilde{n} = |t\tilde{n}$. States are represented by unit vectors and all vectors of a decomposition such as $|u\tilde{n}$ and $|w\tilde{n}$ have unit length, are mutually orthogonal and generate the whole vector space, thus $|a_0|^2 + |a_1|^2 = 1$.

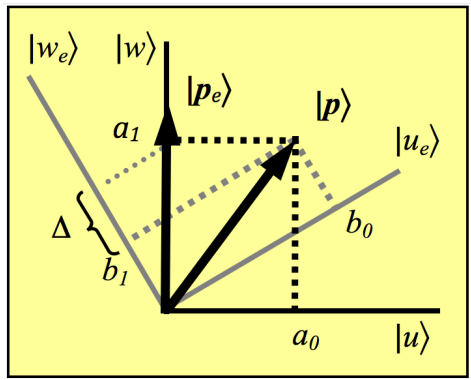

Figure 2. Graphical depiction of a vector $|p\tilde{n}$ representing particular state of TIRE, specifically, a state in which the tread is worn away. In the default context, the state of tire is more likely to collapse to the projection vector $|w\tilde{n}$ which represents wasteful than to its orthogonal projection vector $|u\tilde{n}$ which represents useful. This can be seen by the fact that subspace $a_0$ is smaller than subspace $a_1$. Under the influence of the context playground equipment, the opposite is the case, as shown by the fact that $b_0$ is larger than $b_1$. Also shown is the projection vector after renormalization.

The conception of TIRE changes when activation of the set L of properties of TIRE, e.g.

'weather resistant', spreads to other concepts in the network for which these properties are relevant. Contexts such as <u>playground equipment</u> that share properties with TIRE become candidate members of the set M of relevant contexts for TIRE. The context <u>playground equipment</u>, denoted *e,* consists of the concepts SWING, denoted $|s_e ñ$, and SLIDE, denoted $|l_e ñ$. The restructured conception of TIRE in the context of <u>playground equipment</u>, denoted $|p_e ñ$, is given by $b_0|u_e ñ + b_1|w_e ñ$, where $u_e ñ = b_2|t_e ñ + b_3|t_e s_e ñ + b_4|t_e l_e ñ$, and where $|t_e s_e ñ$ stands for the possibility that a tire functions as a swing, and $|t_e l_e ñ$ stands for the possibility that a tire functions as a slide. The amplitude of $|w_e ñ$, $|b_1|$, is less than $|a_1|$, the amplitude of $|wñ$. This is because $|b_0| > |a_0|$, since $|b_0|$ consists of the possibility of a tire being used not just as a tire, but as a swing or slide. Because certain strongly weighted properties of SLIDE, such as 'long' and 'flat', are not properties of TIRE, $|b_4|$ is small. That is not the case for SWING, so $|b_3|$ is large. Therefore, in the context <u>playground equipment</u>, the concept TIRE has a high probability of collapsing to TIRE SWING, an entangled state of the concepts TIRE and SWING. Entanglement introduces interference of a quantum nature, and hence the amplitudes are complex numbers (Aerts, 2009). If this collapse takes place, TIRE SWING is thereafter a new state of both concepts TIRE and SWING.

This example shows that a formal approach to concept interactions that is consistent with human data (Aerts, 2009; Aerts, Aerts, & Gabora, 2009; Aerts, Broekaert, Gabora, & Veloz, 2012; Aerts, Gabora, & Sozzo, submitted; Hampton, 1987) can model the restructuring of information (*e.g.,* TIRE) under a new context (*e.g.*, playground equipment). Note how in the quantum representation, probability is treated as arising not from a lack of information *per se*, but from the limitations of any particular context (even a 'default' context).

The limitations of this approach are as interesting as its strengths. It is not possible to list, or even develop an algorithm that will list, all possible uses or contexts for any item such as a tire or screwdriver (Longo, Montevil, & Kaufman, 2012). This is what has been referred to as the *frame problem*. As a consequence, human input is particularly welcome at this juncture to define the relevant contexts, *e.g.*, the possible uses of a tire. Studies would be run using data collected from real humans to determine the extent to which the model matches typicality ratings and generation frequencies of exemplars of concepts in particular contexts by human participants, as per (Veloz, Gabora, Eyjolfson, & Aerts, 2011). SCOP models of individual concepts can be embedded into an associative "reaction network". Concept interactions are then modeled as reactions that generate products. Chemical Organization Theory (Dittrich & Speroni di Fenizio, 2008; Dittrich & Winter, 2007; Dittrich, Ziegler, & Banzhaf, 2001), which provides an algebraic means of solving nonlinear, coupled differential equations in reaction networks, or some other such theory, can be used to model the associative structure of interrelated sets of concepts a whole, and study the conditions under which it restores equilibrium in response to the introduction of new states of concepts that results from placing them in new contexts.

Using this SCOP-based cognitive "reaction network" it would be possible to test the hypothesis that contextual focus (the ability to shift between different modes of thought depending on the context– increases cognitive efficiency. If the amplitude associated with $|wñ$ for any concept becomes high—such as for TIRE if the weight of the property 'tread' is low—this signals that the potentiality to re-conceptualize the concept is high. This causes a shift to a more associative mode by increasing α, causing activation of other concepts that share properties with this concept, as described previously.

***Enhanced Computational Model of Cultural Evolution***

Let us now examine how a model of restorative restructuring such as the SCOP-based one we just looked at could be used to develop a cognitively sophisticated computational model of cultural evolution. We will refer to this 'new and improved' model as EVOC2.

So that the EVOC2 agents have something to make artifacts from, their world would contain resource bases from which objects are extracted and wastes are generated. Extracted objects can be joined (lego-style) to construct other objects. Agents have mental representations of resources and objects made from resources. Objects derived from the same resource are modeled in their conceptual networks as states of a concept. Newly extracted or constructed objects have a fitness that defines how useful or wasteful they are *with respect to the other objects an agent has encountered*. Thus existing objects provide contexts that affect the utility of new objects, and an agent's knowledge of existing objects defines its *perspective*.

The artificial culture can now evolve as follows:

*Invent.* Agents invent as in EVOC, except that they invent not actions but objects, using resources in adjacent cells. Extracting an object from a resource creates waste objects.

*Detect and Actualize Potential for Adaptive Change*. If a waste object $p$ is accumulating adjacent to A1, A1 recursively modifies $p$ by considering it from A1's perspective. This continues until $p$ is in a new less wasteful state $p_{A1*}$ which is an eigenstate with respect to A1's perspective. This process may modify not just $p$, but A1's perspective. Perspectives change in response to the ideas and objects an agent interacts with; thus a perspective can encompass more than one context.

*Contextual focus*. The previous step may involve temporarily assuming a more associative processing mode in response to the magnitude of potential for adaptive change.

*Transmission*. Modified object, $p_{A1*}$, becomes input to the associative networks of adjacent agents.

*Context-dependent Restructuring*. If $p_{A1*}$ is wasteful (has potential to change) with respect to the perspective of another agent, A2, then A2 recursively modifies $p_{A1*}$ until it is an eigenstate with respect to A2's perspective, at which point it is referred to as $p_{A1*A2*}$. Since A1's perspective is reflected in $p_{A1*}$, assimilation of $p_{A1*}$ modifies A2's perspective in a way that reflects exposure to (though not necessarily incorporation of or agreement with) A1's perspective. This continues until $p$ settles on stable or cyclic attractor, or we terminate after a set number of iterations (since a chaotic attractor or limit cycle may be hard to distinguish from a non-stable transient).

*Evaluate*. The user assesses the usefulness of the culturally evolved objects for the agents, as well as object diversity, and wastefulness.

EVOC2 will be deemed a success if it not only evolves cultural novelty that is cumulative, adaptive, and open-ended (as in EVOC with RR), but also (a) *restructures* conceptions of objects by viewing them from different perspectives (new contexts), (b) generates inventions that open up niches for other inventions, and (c) exhibits contextual focus, *i.e.,* shifts to an associative mode to restructure and shifts back to fine-tune. It is hypothesized that these features will increase the complexity of economic webs of objects and recycled wastes.

### *Elucidating the Algorithmic Structure of Biological versus Cultural Evolution*

The design features that made EVOC2 specific to the problem of waste recycling can eventually be replaced by general-purpose counterparts, resulting in a *cultural algorithm* (CAL[1]). It will be interesting to **c**ompare the performance of a CAL with a GA on standard problems (*e.g.*, the Rosenbrock function) as well as on insight tasks such as real-world waste recycling webs that require restructuring. Waste recycling is a particularly appropriate application because it explicitly requires considering how the same item offers a different set of constraints and affordances when considered with respect to a different goal, a different demographic, or a different aesthetic sensibility (one person's trash is another person's treasure). In general the CAL is expected to outperform the GA on problems that involve not just multiple *constraints* but multiple *perspectives, e.g.*, economic and environmental.

---

[1] Cultural algorithm is abbreviated CAL because CA customarily refers to cellular automaton.

A long-term objective is to develop an integrated framework for evolutionary processes that encompasses natural selection, cultural evolution, and communal exchange theories of early life. Finally, it can advance knowledge of how systems evolve. Early efforts toward a general cross-disciplinary framework for evolution Processes were modeled as *context-dependent actualization of potential*: an entity has potential to change various ways, and how it *does* change depends on the contexts it interacts with (Gabora & Aerts, 2005, 2007). These efforts focused on distinguishing processes according to the degree of non-determinism they entail, and the extent to which they are sensitive to, internalize, and depend upon a particular context. With the sorts of tools outlined here, it will be possible to compare the effectiveness of communal exchange, Darwinian, and mixed strategies in different environments (simple versus complex, static versus fluctuating, and so forth. This will result in a more precise understanding of the similarities and differences between biological and cultural evolution, and help us recognize other evolutionary processes that we may discover as science penetrates ever deeper into the mysteries of our universe.

## Summary and Conclusions

Culture evolves with breathtaking speed and efficiency. We are crossing the threshold to an exciting frontier: a scientific understanding of the process by which cultural change occurs, as well as the means to capitalize on this understanding. The cultural evolution inspired human computation program of research described in this chapter is ambitious and interdisciplinary, but it builds solidly on previous accomplishments.

We examined evidence that culture evolves through a non-Darwinian communal exchange process, and discussed a plan for modeling the autopoietic structures that evolve through biological and cultural processes—*i.e.*, metabolic reaction networks and associative networks. This will make it possible to undertake a comparative investigation of the dynamics of communally exchanging groups of these two kinds of networks. This research is necessary to achieve a unification of the social and behavioral sciences comparable to Darwin's unification of the life sciences.

Efforts are underway toward the development of a computational model of cultural evolution that incorporates the kind of sophisticated cognitive machinery by which cultural novelty evolves. These include the combining of concepts to give rise to new concepts sometimes with emergent properties, and the capacity to shift between different modes of thought depending on the situation. An important step is to embed formal models of concepts in a modified "reaction network" architecture, in order to computationally model how clusters of interrelated concepts modify one another to achieve a more stable lower energy state, through a process we referred to as *context-driven restorative restructuring*. Efforts are also underway toward the development of a computer program for identifying patterns of historical relationship amongst sets of artifacts. Human input is used to define *contexts*—perspectives or situations that define which features or attributes are potentially relevant. One long-term objective of this kind of research program is to develop a cultural algorithm: an optimization and problem-solving tool inspired by cultural evolution. This will allow us to investigate how strategies for recursively re-processing and restructuring information, or shifting between different processing modes, affect the capacity to evolve cumulative, adaptive, open-ended novelty.

The ideas presented in this chapter are speculative, ambitious, and innovative both conceptually and methodologically, but they have far-reaching implications and potentially diverse applications. The human computation program proposed here could promote a scientific understanding of the current accelerated pace of cultural change and its transformative effects on humans and our planet. It may foster cultural developments that are healthy and productive in the long term as well as the short term, and help us find solutions to complex crises we now face.

## Acknowledgements

This research was conducted with the assistance of grants from the National Science and Engineering Research Council of Canada, and the Fund for Scientific Research of Flanders, Belgium.

Gabora - Cultural Evolution as Distributed Computation  11Dawkins, R. (1976). *The selfish gene.* Oxford: Oxford Univ. Press.
DiPaola, S., & Gabora, L. (2009). Incorporating characteristics of human creativity into an evolutionary art algorithm. *Genetic Programming and Evolvable Machines, 10*(2), 97-110.
Dittrich, P., & Speroni di Fenizio, P. (2008). Chemical organization theory. *Bulletin of Mathematical Biology, 69,* 1199–1231.
Dittrich, P, & Winter, L. 2007. Chemical organizations in a toy model of the political system. *Advances in Complex Systems, 1*(4), 609– 627.
Dittrich, P., Ziegler, J., & Banzhaf, W. 2001. Artificial chemistries – a review. *Artificial Life, 7*(3), 225–275.
Donald, M. (1991). *Origins of the modern mind*, Cambridge, MA: Harvard Univ. Press.
Durham, W. (1991). *Coevolution: Genes, culture, and human diversity.* Stanford: Stanford Univ. Press.
Eliasmith, C., & Thagard, P. (2001). Integrating structure and meaning: A distributed model of analogical mapping. *Cognitive Science, 25,* 245–286.
Fracchia J. & Lewontin, R. C. (1999). Does culture evolve? *History & Theory,* 38, 52–78.
Gabora, L. (1995). Meme and variations: A computer model of cultural evolution. In (L. Nadel & D. Stein, Eds.) *1993 Lectures in complex systems* (pp. 471–486). Boston: Addison-Wesley.
Gabora, L. (1996). A day in the life of a meme. *Philosophica, 57,* 901-938.
Gabora, L. (1997). The origin and evolution of culture and creativity. *Journal of Memetics: Evolutionary Models of Information Transmission, 1*(1).
Gabora, L. (1998). Autocatalytic closure in a cognitive system: A tentative scenario for the origin of culture. *Psycoloquy, 9*(67). [adap-org/9901002]
Gabora, L. (2003). Contextual focus: A cognitive explanation for the cultural transition of the Middle/Upper Paleolithic. In (R. Alterman & D. Hirsch, Eds.) *Proc Annual Meeting of the Cognitive Science Society* (432-437)*,* Boston MA, 31 July - 2 August. Lawrence Erlbaum.
Gabora, L. (2006a). The fate of evolutionary archaeology: Survival or extinction? *World Archaeology*, *38*(4), 690–696.
Gabora, L. (2006b). Self-other organization: Why early life did not evolve through natural selection. *Journal of Theoretical Biology*, *241*(3), 443–250.
Gabora, L. (2008). The cultural evolution of socially situated cognition. *Cognitive Systems Research*, *9*(1), 104-113.
Gabora, L. (2011). Five clarifications about cultural evolution. *J Cognition and Culture, 11,* 61-83.
Gabora, L., & Aerts, D. (2002). Contextualizing concepts. *Proc 15th Int FLAIRS Conference* (Special Track 'Categorization and Concept Representation: Models and Implications') (pp. 148-152), Pensacola FL, May 14-17, American Association for Artificial Intelligence.
Gabora, L., & Aerts, D. (2002b). Contextualizing concepts using a mathematical generalization of the quantum formalism. *J Exp and Theor Artificial Intelligence, 14*(4), 327–358.
Gabora, L., & Aerts, D. (2005). Evolution as context-driven actualization of potential: Toward an interdisciplinary theory of change of state. *Interdisc Science Reviews*, *30*(1), 69–88.
Gabora, L., & DiPaola, S. (2012). How did humans become so creative? *Proceedings of the International Conf on Computational Creativity* (pp. 203-210). May 31 - June 1, Dublin.
Gabora, L., Leijnen, S., Veloz, T., & Lipo, C. (2011). A non-phylogenetic conceptual network architecture for organizing classes of material artifacts into cultural lineages. *Proc Ann Mtng Cog Sci Soc*. July 20-23, 2011, Boston MA.
Gabora, L., O'Connor, B., & Ranjan, A. (2012). The recognizability of individual creative styles within and across domains. *Psychology of Aesthetics, Creativity, and the Arts*, *6*(4), 351-360.

Gabora - Cultural Evolution as Distributed Computation    13Tëmkin, I., & Eldredge, N. (2007). Phylogenetics and material cultural evolution. *Current Anthropology, 48,* 146−153.

Thagard, P., & Stewart, T. C. (2011). The AHA! experience: Creativity through emergent binding in neural networks. *Cognitive Science, 35,* 1–33.

Veloz, T., Gabora, L., Eyjolfson, M., & Aerts, D. (2011). A model of the shifting relationship between concepts and contexts in different modes of thought. *Proceedings of the Fifth International Symposium on Quantum Interaction*. June 27, 2011, Aberdeen UK.

Veloz, T., Tëmkin, I., & Gabora, L., A conceptual network-based approach to inferring cultural phylogenies. Proceedings Annual Meeting Cognitive Science Society. Sapporo Japan, 2012.

Vetsigian, K., Woese, C., & Goldenfeld, N. (2006). Collective evolution and the genetic code. *Proceedings of the National Academy of Science, 103,* 10696–10701.

Williams, R. J. P., & Frausto da Silva, J. J. R. (2003). Evolution was chemically constrained. *Journal of Theoretical Biology, 220,* 323–343.

Woese, C. R. (2002). On the evolution of cells. *Proceedings of the National Academy of Science, 99,* 8742–8747.